\documentclass[conference]{IEEEtran}%
\usepackage{amsfonts}
\usepackage{amsmath}
\usepackage{amssymb}
\usepackage{graphicx}%
\usepackage{braket}
\usepackage{hyperref}
\providecommand{\U}[1]{\protect\rule{.1in}{.1in}}
\newtheorem{theorem}{Theorem}

\newtheorem{remark}[theorem]{Remark}

\begin{document}

\title{Quantum polar codes for arbitrary channels}
\author{\IEEEauthorblockN{Mark M.\ Wilde} \IEEEauthorblockA{\it School of Computer Science, McGill University\\Montreal, Quebec, Canada}\and \IEEEauthorblockN{Joseph M.\ Renes} \IEEEauthorblockA{\it Institut f\"ur Theoretische Physik, ETH Zurich\\Z\"urich, Switzerland}}
\maketitle
\begin{abstract}
We construct a new entanglement-assisted quantum polar coding scheme which achieves the symmetric coherent information rate by synthesizing 
\textquotedblleft amplitude\textquotedblright\ and \textquotedblleft
phase\textquotedblright\ channels from a given, \emph{arbitrary} quantum channel. 
We first demonstrate the coding scheme for arbitrary quantum channels with qubit inputs,
and we show that quantum data can be reliably decoded by  
$O(N)$ rounds of coherent quantum successive cancellation, followed by
$N$ controlled-NOT gates (where $N$ is the number of
channel uses). We also find that the 
entanglement consumption rate of the code vanishes for \emph{degradable} quantum
 channels. Finally, we extend the coding scheme to channels with multiple qubit inputs. This gives  
a near-explicit method for realizing one of the most striking phenomena in
quantum information theory: the \textit{superactivation effect}, whereby two
quantum channels which individually have zero quantum capacity can have a
non-zero quantum capacity when used together.

\end{abstract}

Polar coding is a promising code construction for transmitting classical
information over classical channels \cite{A09}. Arikan proved that polar codes
achieve the symmetric capacity of any classical channel \cite{A09}, with an
encoding and decoding complexity that is $O\left(  N\log N\right)  $ where $N$
is the number of channel uses. These codes exploit the channel polarization
effect whereby a particular recursive encoding induces a set of virtual
channels, such that a fraction of the virtual channels are perfect for data
transmission while the other fraction are useless for this task. The fraction
containing perfect virtual channels is equal to the channel's symmetric
capacity.

In this paper, we offer a new quantum polar coding scheme strongly based on
ideas of Renes and Boileau \cite{RB08}, who showed that quantum coding protocols can be constructed from two different protocols
that protect classical information encoded into complementary observables. In
particular, a protocol for reliably transmitting quantum data can be built
from a protocol that reliably recovers classical information encoded into an
\textquotedblleft amplitude\textquotedblright\ observable and a protocol that
reliably recovers \textquotedblleft phase\textquotedblright\ information with
the assistance of quantum side information (see
Refs.~\cite{RB09,BR09,R10,Renes08062011}\ for related ideas). 

These ideas were 
used to construct a quantum polar coding scheme with an efficient decoder in~\cite{RDR11}, but only for a certain set of channels with essentially classical outputs.
Following a different approach, Ref.~\cite{WG11a} constructed quantum polar codes for degradable channels. Our new quantum polar
coding scheme has several advantages over these previous schemes:

\begin{itemize}
\item The net rate of quantum communication is equal to the symmetric coherent
information for an \textit{arbitrary} quantum channel with qubit input.

\item The decoder is \textit{explicit}, and consists of $O\left(  N\right)
$\ rounds of coherent quantum successive cancellation followed by $N$ CNOT gates.

\item The entanglement consumption rate vanishes for an \textit{arbitrary}
degradable channel with qubit input.
\end{itemize}

Following the multi-level coding method of Ref.~\cite{STA09},
we show how to extend the coding scheme to channels with multiple qubit inputs.
This gives an explicit code construction for the superactivation effect, in
which two zero-capacity channels have a
non-zero quantum capacity when used together \cite{SY08} (in this sense, the
channels \textit{activate} each other). 


\section{Quantum Polar Coding Scheme}

\subsection{Classical-quantum channels for complementary variables}

Consider a quantum channel $\mathcal{N}$\ with a two-dimensional input system
$A^{\prime}$ and a $d$-dimensional output system $B$. Let $U_{\mathcal{N}%
}^{A^{\prime}\rightarrow BE}$ denote the isometric extension of this channel.
Let $\left\vert z\right\rangle $ denote the computational or \textquotedblleft
amplitude\textquotedblright\ basis with $z\in\left\{  0,1\right\}  $, and let
$\left\vert \widetilde{x}\right\rangle $ denote the conjugate, Hadamard, or
\textquotedblleft phase\textquotedblright\ basis with $\widetilde{x}%
\in\left\{  +,-\right\}  $ and $\left\vert \pm\right\rangle
\equiv\left(  \left\vert 0\right\rangle \pm\left\vert 1\right\rangle \right)
/\sqrt{2}$.

Following Ref.~\cite{RB08}, we consider building up a quantum communication
protocol from two classical communication protocols that preserve classical
information encoded into complementary variables. In this vein, two particular
classical-quantum (cq) channels are important. First, consider the cq channel
induced by sending an amplitude basis state over  $\mathcal{N}$:%
\begin{equation}
W_{A}:z\rightarrow\mathcal{N}^{A^{\prime}\rightarrow B}\left(  \left\vert
z\right\rangle \left\langle z\right\vert \right)  \equiv\phi^B_{z}%
,\label{eq:amplitude-cq-channel}%
\end{equation}
where the classical input $z$ is a binary variable and the notation $W_{A}$
indicates that the classical information is encoded into the amplitude basis. We can 
regard this as the sender (Alice) modulating a standard signal $\ket{0}$ with $X^z$ and transmitting the 
result to the receiver (Bob). 

For the other cq channel, suppose that Alice instead transmits a binary variable $x$ by modulating the signal with $Z^x$, a 
rephasing of the amplitude basis states. However, instead of applying this to $\ket{0}$, she modulates one half of an entangled qubit pair (ebit) shared with Bob. These qubits are in the state
\[
\left\vert \Phi\right\rangle ^{CA^{\prime}}\equiv\frac{1}{\sqrt{2}}\sum
_{z\in\left\{  0,1\right\}  }\left\vert z\right\rangle ^{C}\left\vert
z\right\rangle ^{A^{\prime}}=\frac{1}{\sqrt{2}}\sum_{\widetilde{x}\in\left\{
+,-\right\}  }\left\vert \widetilde{x}\right\rangle ^{C}\left\vert
\widetilde{x}\right\rangle ^{A^{\prime}},
\]
with Alice holding $A'$ and Bob $C$. 
The modulation yields
\begin{align}
\ket{\sigma_x}^{BCE}&=U_{\mathcal{N}%
}^{A^{\prime}\rightarrow BE} \left(Z^x\right)^{A'}\ket{\Phi}^{A'C},\\
&=\tfrac1{\sqrt{2}}\sum_{z\in\{0,1\}}(-1)^{xz}\ket{\phi_z}^{BE}\ket{z}^C,
\end{align} 
where $\ket{\phi_z}^{BE}$ is a purification of $\phi_z^{B}$ in (\ref{eq:amplitude-cq-channel}). The resulting cq channel is then of the following form:%
\begin{equation}
W_{P}:x\rightarrow\sigma_{x}^{BC},\label{eq:phase-cq-channel}%
\end{equation}
where the notation $W_{P}$
indicates that the classical information is encoded into a phase variable. 
In contrast to $W_A$, the channel $W_{P}$ is one in which the receiver has quantum side information
(in the form of system $C$)
that is helpful for decoding the transmitted phase
information.\footnote{Operationally, this quantum side information becomes
available to Bob after he coherently decodes the amplitude variable. It does
\textit{not} correspond operationally to a Bell state shared before
communication begins.} 

Both cq channels in (\ref{eq:amplitude-cq-channel}) and
(\ref{eq:phase-cq-channel}) arise in the error analysis of our quantum polar
coding scheme, in the sense that its performance depends on the performance of
constituent polar codes constructed for these cq channels. Moreover, the two channels
are more closely related than they may initially appear. To see their relationship, consider the state
\begin{align*}
\label{eq:channelstate}
\ket{\psi}&=\tfrac1{\sqrt{2}}\hspace{-2mm}\sum_{x\in\{0,1\}}\ket{\widetilde{x}}^A\ket{\sigma_x}^{BCE}=\tfrac1{\sqrt{2}}\hspace{-2mm}\sum_{z\in\{0,1\}}\ket{z}^A\ket{z}^C\ket{\phi_z}^{BE}.
\end{align*} 
Measuring system $A$ in the phase basis $\ket{\widetilde{x}}$ generates the $W_P$ output state $\sigma_x^{BE}$, while measuring $A$ in the amplitude basis generates the $W_A$ output $\phi_z^B$.  

Another important channel is the cq channel $W_{E}$ induced to the environment
when inputting amplitude-encoded classical information: $W_{E}:z\rightarrow
{\rm Tr}_{B}\{U_{\mathcal{N}}^{A^{\prime}\rightarrow BE}\left(  \left\vert
z\right\rangle \left\langle z\right\vert \right)  \}$. We do not consider this
channel for our quantum polar coding scheme or its error analysis, but we
instead consider it in Section~\ref{sec:ent-cons-rate-vanish}\ when relating
the quantum polar coding scheme of this paper to the previous one from
Ref.~\cite{WG11a}.

\subsection{Channel Polarization}

Two channel parameters that determine the performance of a cq
channel $W:x\rightarrow\rho_{x}$ are the fidelity $F\left(  W\right)
\equiv\left\Vert \sqrt{\rho_{0}}\sqrt{\rho_{1}}\right\Vert _{1}^{2}$ and the
symmetric Holevo information $I\left(  W\right)  \equiv H\left(  \left(
\rho_{0}+\rho_{1}\right)  /2\right)  -\left[  H\left(  \rho_{0}\right)
+H\left(  \rho_{1}\right)  \right]  /2$ where $H\left(  \sigma\right)
\equiv-$Tr$\left\{  \sigma\log_{2}\sigma\right\}  $ is the von Neumann
entropy. These parameters generalize the Bhattacharya parameter and the
symmetric mutual information \cite{A09}, respectively, and are related as
$I\left(  W\right)  \approx1\Leftrightarrow F\left(  W\right)  \approx0$ and
$I\left(  W\right)  \approx0\Leftrightarrow F\left(  W\right)  \approx1$
\cite{WG11}. The channel $W$ is near perfect when $I\left(  W\right)
\approx1$ and near useless when $I\left(  W\right)  \approx0$.

Ref.~\cite{WG11} demonstrated how to construct synthesized
versions of  $W$, by channel combining and splitting \cite{A09}. For blocksize $N$, the synthesized channels are of the following
form:%
\begin{equation}
W_{N}^{\left(  i\right)  }:u_{i}\rightarrow\rho_{\left(  i\right)  ,u_{i}%
}^{U_{1}^{i-1}B^{N}},\label{eq:split-channels}%
\end{equation}
where%
\begin{align}
\rho_{\left(  i\right)  ,u_{i}}^{U_{1}^{i-1}B^{N}} &  \equiv\sum_{u_{1}^{i-1}%
}\frac{1}{2^{i-1}}\left\vert u_{1}^{i-1}\right\rangle \left\langle u_{1}%
^{i-1}\right\vert ^{U_{1}^{i-1}}\otimes\overline{\rho}_{u_{1}^{i}}^{B^{N}},\\
\overline{\rho}_{u_{1}^{i}}^{B^{N}} &  \equiv\sum_{u_{i+1}^{N}}\frac
{1}{2^{N-i}}\rho_{u^{N}G_{N}}^{B^{N}},\,\,\,\,\,\,\,\,
\rho^{B^N}_{x^{N}}   \equiv\rho^{B_1}_{x_{1}}\otimes\cdots\otimes\rho^{B_N}_{x_{N}}, \nonumber
\end{align}
and $G_{N}$ is Arikan's encoding circuit matrix built from classical CNOT\ and
permutation gates. The interpretation of this channel is that it is the one
\textquotedblleft seen\textquotedblright\ by the input $u_{i}$ if all of the
previous bits $u_{1}^{i-1}$ are available and if we consider all the future
bits $u_{i+1}^{N}$ as randomized. This motivates the development of a quantum
successive cancellation decoder (QSCD) \cite{WG11}\ that attempts to distinguish
$u_{i}=0$ from $u_{i}=1$ by adaptively exploiting the results of previous
measurements and quantum hypothesis tests for each bit decision.

The synthesized channels $W_{N}^{\left(  i\right)  }$ polarize, in the sense
that some become nearly perfect for classical data transmission while others
become nearly useless. To prove this result, one can model the
channel splitting and combining process as a random birth process
\cite{A09,WG11}, and one can demonstrate that the induced random birth
processes corresponding to the channel parameters $I(W_{N}^{\left(  i\right)
})$ and $F(W_{N}^{\left(  i\right)  })$ are martingales that converge almost
surely to zero-one valued random variables in the limit of many recursions.
The following theorem characterizes the rate with which the channel
polarization effect takes hold \cite{WG11}, and it is useful in proving
statements about the performance of polar codes for cq channels:

\begin{theorem}
\label{thm:fraction-good}Given a binary input cq channel $W$ and any
$\beta<1/2$, it holds that $\lim_{n\rightarrow\infty}\Pr_{I}\{\sqrt
{F(W_{2^{n}}^{\left(  I\right)  })}<2^{-2^{n\beta}}\}=I\left(  W\right)  $,
where $n$ indicates the level of recursion for the encoding, $W_{2^{n}%
}^{\left(  I\right)  }$ is a random variable characterizing the $I^{\text{th}%
}$ split channel, and $F(W_{2^{n}}^{\left(  I\right)  })$ is the fidelity of
that channel.
\end{theorem}

Assuming knowledge of the good and bad channels, one can then construct a
coding scheme based on the channel polarization effect, by dividing the
synthesized channels according to the following polar coding rule:%
\begin{equation}
\mathcal{G}_{N}\left(  W,\beta\right)  \equiv\left\{  i\in\left[  N\right]
:\sqrt{F(W_{N}^{\left(  i\right)  })}<2^{-N^{\beta}}\right\}  ,
\label{eq:polar-coding-rule}%
\end{equation}
and $\mathcal{B}_{N}\left(  W,\beta\right)  \equiv\left[  N\right]
\setminus\mathcal{G}_{N}\left(  W,\beta\right)  $, so that $\mathcal{G}%
_{N}\left(  W,\beta\right)  $ is the set of \textquotedblleft
good\textquotedblright\ channels and $\mathcal{B}_{N}\left(  W,\beta\right)  $
is the set of \textquotedblleft bad\textquotedblright\ channels. The sender
then transmits the information bits through the good channels and
\textquotedblleft frozen\textquotedblright\ bits through the bad ones. A
helpful assumption for error analysis is that the frozen bits are chosen
uniformly at random such that the sender and receiver both have access to
these frozen bits. Ref.~\cite{WG11} provided an explicit construction of a
QSCD that has an error probability equal to
$o(2^{-N^{\beta}})$---let $\{\Lambda_{u_{\mathcal{A}}}^{\left(  u_{\mathcal{A}%
^{c}}\right)  }\}$ denote the corresponding decoding POVM, with
$u_{\mathcal{A}}$ the information bits and $u_{\mathcal{A}^{c}}$ the frozen bits.

For our quantum polar coding scheme, we exploit a coherent version of Arikan's
encoder \cite{A09}, meaning that the gates are quantum CNOTs and permutations
(this is the same encoder as in Refs.~\cite{RDR11,WG11a}). When sending
amplitude-basis classical information through the encoder and
channels, the effect is to induce synthesized channels $W_{A,N}^{\left(
i\right)  }$ as described above. Theorem~\ref{thm:fraction-good}\ states that
the fraction of amplitude-good channels (according to the criterion in
(\ref{eq:polar-coding-rule})) is equal to $I\left(  Z;B\right)_\phi  $ where the
Holevo information $I\left(  Z;B\right)_\phi $ is evaluated with respect to the cq
state $\phi^{ZB}=\frac{1}{2}\sum_{z\in\left\{  0,1\right\}  }\left\vert
z\right\rangle \left\langle z\right\vert ^{Z}\otimes\phi_{z}^{B}$, with
$\phi_{z}^B$ defined in (\ref{eq:amplitude-cq-channel}). It will be convenient to express this 
quantity as $I(Z^A;B)_\psi$ using the state $\ket{\psi}$, 
where the $Z^A$ indicates that system $A$ is first measured in the amplitude basis.

As in~\cite{RDR11}, the same encoding operation leads to channel polarization for the phase channel $W_P$ as well. 
Suppose Alice modulates her halves of the entangled pairs as before, but then inputs them to the coherent encoder before
sending them via the channel to Bob. The result is  
\begin{align}
\label{eq:phaseencstate}
\frac1{\sqrt{2^N}}\sum_{z^N\in \{0,1\}^N}(-1)^{x^N\cdot z^N}\ket{\phi_{z^NG_N}}^{B^NE^N}\ket{z^N}^{C^N},
\end{align}
whose $B^NC^N$ marginal state is simply $U_{\mathcal{E}}^{C^N}\sigma_{x^N G_N^T}^{B^NC^N}U_{\mathcal{E}}^{\dagger C^N}$, where $U_{\mathcal{E}}$ denotes the polar encoder. Here we have used the fact that the matrix corresponding to $G_N$ is invertible. Thus, the coherent encoder also induces synthesized channels $W_{P,N}^{(i)}$ using the encoding matrix $G_N^T$ instead of $G_N$, modulo the additional $U_{\mathcal{E}}$ acting on $C^N$. 
Note that the classical side information for the
$W_{P,N}^{\left(  i\right)  }$ is different from that in (\ref{eq:split-channels})
 because the direction of all CNOT\ gates is flipped due to the transpose of $G_N$
when acting on phase variables. The change in the direction of the CNOT\ gates
means that the $i^{\text{th}}$ synthesized phase channel $W_{P,N}^{\left(
i\right)  }$ is such that all of the \textit{future} bits $x_{N}\cdots
x_{i+1}$ are available to help in decoding bit $x_{i}$ while all of the
\textit{previous} bits $x_{i-1}\cdots x_{1}$ are randomized. (This is the same
as described in Ref.~\cite{RDR11}\ for Pauli channels.) 

For the channel in
(\ref{eq:phase-cq-channel}), the fraction of phase-good channels is
approximately equal to $I\left(  X;BC\right)_\sigma  $, where the Holevo information
$I\left(  X;BC\right)_\sigma  $ is with respect to a cq state of the form
$\frac{1}{2}\sum_{x\in\left\{  0,1\right\}  }\left\vert x\right\rangle
\left\langle x\right\vert ^{X}\otimes\sigma_{x}^{BC}$, with $\sigma_{x}^{BC}$
in (\ref{eq:phase-cq-channel}). Again, we can formulate this using $\ket{\psi}$ as $I(X^A;BC)_\psi$, 
this time $X^A$ indicating $A$ is measured in the phase basis.

Lemma~2 of Ref.~\cite{RB08}\ outlines an important relationship between the
Holevo information of the phase channel to Bob and the Holevo information of
the amplitude channel to Eve, which for our case reduces to $I\left(
X^A;BC\right)_\psi  =1-I\left(  Z^A;E\right)_\psi  $. This relationship already suggests
that channels which are phase-good for Bob should be amplitude-bad for Eve and
that channels which are amplitude-good for Eve should be phase-bad for Bob,
allowing us in Section~\ref{sec:ent-cons-rate-vanish} to relate the present
quantum polar coding scheme to that from Ref.~\cite{WG11a}.

\subsection{Coding scheme}

We divide the synthesized cq amplitude channels $W_{A,N}^{\left(  i\right)  }$
into sets $\mathcal{G}_{N}\left(  W_{A},\beta\right)  $ and $\mathcal{B}%
_{N}\left(  W_{A},\beta\right)  $ according to (\ref{eq:polar-coding-rule}),
and similarly, we divide the synthesized cq phase channels $W_{P,N}^{\left(
i\right)  }$ into sets $\mathcal{G}_{N}\left(  W_{P},\beta\right)  $ and
$\mathcal{B}_{N}\left(  W_{P},\beta\right)  $, where $\beta<1/2$. The synthesized
channels correspond to particular inputs to the encoding operation, and thus the
set of all inputs divides into four groups:\ those 
that are good for both the amplitude and phase variable, those that are good
for amplitude and bad for phase, bad for amplitude and good for phase, and
those that are bad for both variables. We establish notation for these
channels as follows:%
\begin{align*}
\mathcal{A} &  \equiv\mathcal{G}_{N}\left(  W_{A},\beta\right)  \cap
\mathcal{G}_{N}\left(  W_{P},\beta\right)  ,\\
\mathcal{X} &  \equiv\mathcal{G}_{N}\left(  W_{A},\beta\right)  \cap
\mathcal{B}_{N}\left(  W_{P},\beta\right)  ,
\end{align*}
\begin{align*}
\mathcal{Z} &  \equiv\mathcal{B}_{N}\left(  W_{A},\beta\right)  \cap
\mathcal{G}_{N}\left(  W_{P},\beta\right)  ,\\
\mathcal{B} &  \equiv\mathcal{B}_{N}\left(  W_{A},\beta\right)  \cap
\mathcal{B}_{N}\left(  W_{P},\beta\right)  .
\end{align*}
Our quantum polar coding scheme has the sender transmit information qubits
through the inputs in $\mathcal{A}$, frozen bits in the phase basis through
the inputs in $\mathcal{X}$, frozen bits in the amplitude basis through the
inputs in $\mathcal{Z}$, and halves of ebits \cite{BDH06} through the inputs in
$\mathcal{B}$ (we can think of these in some sense as being frozen
simultaneously in both the amplitude and phase basis). It is straightforward
to prove (see Appendix~\ref{app:info-set-size}) that the net rate \cite{BDH06} of quantum
communication $\left(  \left\vert \mathcal{A}\right\vert -\left\vert
\mathcal{B}\right\vert \right)  /N$ is equal to the coherent information
$I\left(  A\rangle B\right)  \equiv H\left(  B\right)  -H\left(  AB\right)  $
by observing that the fraction of amplitude-good channels is $I\left(
Z^A;B\right)_\psi  $, the fraction of phase-good channels is $I\left(  X^A;BC\right)_\psi
$, and exploiting the relation $I\left(  X^A;BC\right)_\psi  =1-I\left(  Z^A;E\right)_\psi
$.

\subsection{Error Analysis}

\label{sec:err-analysis}We now demonstrate that this coding scheme works well.
The sender and receiver begin with the following state:%
\begin{align*}
\ket{\Psi_0}=
N_{0}\sum_{u_{\mathcal{A}},u_{\mathcal{B}}}\left\vert u_{\mathcal{A}%
}\right\rangle \left\vert u_{\mathcal{A}}\right\rangle \left\vert
u_{\mathcal{Z}}\right\rangle \left\vert \widetilde{u}_{\mathcal{X}}\right\rangle
\left\vert u_{\mathcal{B}}\right\rangle\otimes \left\vert u_{\mathcal{B}%
}\right\rangle ,
\end{align*}
where Alice possesses the first five registers, Bob
the last one,\footnote{In quantum information theory
 the tensor product symbol is often used implicitly. Our convention  is to leave
it implicit for systems belonging to the same party and use it explicitly 
to denote a division between two parties.} and $N_{0}\equiv1/\sqrt{2^{\left\vert \mathcal{A}\right\vert
+\left\vert \mathcal{B}\right\vert }}$. We also assume for now that the bits
in $u_{\mathcal{Z}}$ and $u_{\mathcal{X}}$ are chosen uniformly at random and
are known to both the sender and receiver. Note that the $4^{\text{th}}$ register is expressed in the phase basis; the amplitude 
basis instead gives 
\begin{align*}
\ket{\Psi_0}=
N_{1}\hspace{-3mm}\sum_{u_{\mathcal{A}},u_{\mathcal{B}},v_{\mathcal{X}}}\hspace{-3mm}\left(  -1\right)
^{u_{\mathcal{X}}\cdot v_{\mathcal{X}}}\!\left\vert u_{\mathcal{A}}\right\rangle
\left\vert u_{\mathcal{A}}\right\rangle \left\vert u_{\mathcal{Z}%
}\right\rangle \left\vert v_{\mathcal{X}}\right\rangle \left\vert
u_{\mathcal{B}}\right\rangle \otimes \left\vert u_{\mathcal{B}}\right\rangle ,
\end{align*}
where 
 $N_{1}\equiv
1/\sqrt{2^{\left\vert \mathcal{A}\right\vert +\left\vert \mathcal{B}%
\right\vert +\left\vert \mathcal{X}\right\vert }}$. The sender then feeds the
middle four registers through the polar encoder and channel, leading to a state of the following form:%
\begin{align*}
\ket{\Psi_1}=
N_{1}\hspace{-3mm}\sum_{u_{\mathcal{A}},u_{\mathcal{B}},v_{\mathcal{X}}}\hspace{-3mm}\left(  -1\right)
^{u_{\mathcal{X}}\cdot v_{\mathcal{X}}}\left\vert u_{\mathcal{A}}\right\rangle\otimes
\left\vert \phi_{u_{\mathcal{A}},u_{\mathcal{Z}},v_{\mathcal{X}}%
,u_{\mathcal{B}}}\right\rangle ^{B^NE^N}\!\!\!\left\vert u_{\mathcal{B}%
}\right\rangle ,
\end{align*}
where $\left\vert \phi_{u_{\mathcal{A}},u_{\mathcal{Z}},v_{\mathcal{X}%
},u_{\mathcal{B}}}\right\rangle ^{B^NE^N}\equiv U_{\mathcal{N}%
}^{\otimes N}U_{\mathcal{E}%
}\left\vert u_{\mathcal{A}}\right\rangle \left\vert u_{\mathcal{Z}%
}\right\rangle \left\vert v_{\mathcal{X}}\right\rangle \left\vert
u_{\mathcal{B}}\right\rangle $ (abusing notation, the encoding operation $G_N$ is left implicit).

Observe that, conditioned on amplitude measurements of $\ket{u_{\mathcal{A}}}$ and $\ket{u_{\mathcal{B}}}$, the $B^N$ subsystem is identical to the polar-encoded
output of $W_A$. Thus, the first step of
the decoder is the following coherent implementation of the QSCD for $W_A$ as in
(\ref{eq:amplitude-cq-channel}):%
\begin{equation}
\sum_{u_{\mathcal{A}},u_{\mathcal{B}},v_{\mathcal{X}}}\sqrt{\Lambda
_{u_{\mathcal{A}},v_{\mathcal{X}}}^{\left(  u_{\mathcal{B}},u_{\mathcal{Z}%
}\right)  }}\otimes\left\vert u_{\mathcal{A}}\right\rangle \left\vert
v_{\mathcal{X}}\right\rangle \otimes\left\vert u_{\mathcal{B}}\right\rangle
\left\vert u_{\mathcal{B}}\right\rangle \left\langle u_{\mathcal{B}%
}\right\vert \otimes\left\vert u_{\mathcal{Z}}\right\rangle
.\label{eq:amp-decoder}%
\end{equation}
The idea here is that the decoder is coherently recovering the bits in
$u_{\mathcal{A}}$ and $v_{\mathcal{X}}$ while using those in $u_{\mathcal{Z}}$
and $u_{\mathcal{B}}$ as classical and quantum side information, respectively.
After doing so, the resulting state is $o(2^{-N^{\beta}})$-close in expected
trace distance to the following ideal state (see
Appendix~\ref{app:err-analysis}):%
\begin{multline*}
\ket{\Psi_2}=N_{1}\hspace{-2mm}\sum_{u_{\mathcal{A}},u_{\mathcal{B}},v_{\mathcal{X}}}\hspace{-2mm}\left(  -1\right)
^{u_{\mathcal{X}}\cdot v_{\mathcal{X}}}\left\vert u_{\mathcal{A}}\right\rangle
\left\vert \phi_{u_{\mathcal{A}},u_{\mathcal{Z}},v_{\mathcal{X}}%
,u_{\mathcal{B}}}\right\rangle ^{B^{N}E^{N}}\otimes\\
\left\vert u_{\mathcal{A}}\right\rangle \left\vert v_{\mathcal{X}%
}\right\rangle \left\vert u_{\mathcal{B}}\right\rangle \left\vert
u_{\mathcal{B}}\right\rangle \left\vert u_{\mathcal{Z}}\right\rangle .
\end{multline*}
The expectation is with respect to the uniformly random choice of 
$u_{\mathcal{X}}$. Thus, Bob has coherently recovered the bits
$u_{\mathcal{A}}$ and $v_{\mathcal{X}}$ with the decoder in
(\ref{eq:amp-decoder}), while making a second coherent and incoherent copy of
the bits $u_{\mathcal{B}}$ and $u_{\mathcal{Z}}$, respectively.

The next step in the process is to make coherent use of the $W_P$ decoder. For this to be useful, however, 
we must show that encoded versions of $\ket{\sigma_x}^{BCE}$, as in (\ref{eq:phaseencstate}), are present in $\ket{\Psi_2}$. To see this, first observe that we can write 
\begin{multline*}
\ket{\Psi_2}=
N_{2}\sum_{\substack{u_{\mathcal{A}},u_{\mathcal{B}},v_{\mathcal{X}%
},\\{x}_{\mathcal{A}},{x}_{\mathcal{B}}}}\left(
-1\right)  ^{u_{\mathcal{X}}\cdot v_{\mathcal{X}}+{x}_{\mathcal{A}%
}\cdot u_{\mathcal{A}}+{x}_{\mathcal{B}}\cdot u_{\mathcal{B}}%
}\left\vert \widetilde{x}_{\mathcal{A}}\right\rangle \otimes\\
\left\vert \phi_{u_{\mathcal{A}},u_{\mathcal{Z}},v_{\mathcal{X}}%
,u_{\mathcal{B}}}\right\rangle ^{B^{N}E^{N}}\left\vert u_{\mathcal{A}%
}\right\rangle \left\vert v_{\mathcal{X}}\right\rangle \left\vert
u_{\mathcal{B}}\right\rangle \left\vert \widetilde{x}_{\mathcal{B}%
}\right\rangle \left\vert u_{\mathcal{Z}}\right\rangle ,
\end{multline*}
where $N_{2}\equiv1/\sqrt{2^{2\left\vert \mathcal{A}\right\vert +2\left\vert
\mathcal{B}\right\vert +\left\vert \mathcal{X}\right\vert }}$, by expressing the first register and the second $\left\vert u_{\mathcal{B}%
}\right\rangle $ register in the phase basis. 
This is nearly the expression we are looking for, as all the desired phase factors are present, except one corresponding to  $\ket{u_{\mathcal{Z}}}$. 

As $u_{\mathcal{Z}}$ is chosen at random, we can describe it quantum-mechanically as arising from part of an entangled state. The other part is shared by Alice and an inaccessible reference. Including this purification degree of freedom, $\ket{\Psi_2}$ becomes
\begin{align*}
\ket{\Psi_2'}=N_{3}\sum_{\substack{u_{\mathcal{A}},u_{\mathcal{B}},v_{\mathcal{X}%
},\\u_{\mathcal{Z}},{x}_{\mathcal{A}},{x}_{\mathcal{B}}}}\left(
-1\right)  ^{u_{\mathcal{X}}\cdot v_{\mathcal{X}}+{x}_{\mathcal{A}%
}\cdot u_{\mathcal{A}}+{x}_{\mathcal{B}}\cdot u_{\mathcal{B}}%
}\left\vert \widetilde{x}_{\mathcal{A}}\right\rangle \otimes\\
\left\vert \phi_{u_{\mathcal{A}},u_{\mathcal{Z}},v_{\mathcal{X}}%
,u_{\mathcal{B}}}\right\rangle ^{B^{N}E^{N}}\left\vert u_{\mathcal{A}%
}\right\rangle \left\vert v_{\mathcal{X}}\right\rangle \left\vert
u_{\mathcal{B}}\right\rangle \left\vert \widetilde{x}_{\mathcal{B}%
}\right\rangle \left\vert u_{\mathcal{Z}}\right\rangle\otimes \ket{u_{\mathcal{Z}}} ,
\end{align*}
where $N_3=N_2/\sqrt{2^{|\mathcal{Z}|}}$. Again utilizing the phase basis gives 
\begin{align*}
\ket{\Psi_2'}=N_{3}\hspace{-3mm}\sum_{\substack{u_{\mathcal{A}},u_{\mathcal{B}},v_{\mathcal{X}%
},u_{\mathcal{Z}},\\{x}_{\mathcal{A}},{x}_{\mathcal{B}},{x}_{\mathcal{Z}}}}\hspace{-3mm}\left(
-1\right)  ^{u_{\mathcal{X}}\cdot v_{\mathcal{X}}+{x}_{\mathcal{A}%
}\cdot u_{\mathcal{A}}+{x}_{\mathcal{B}}\cdot u_{\mathcal{B}}+{x}_{\mathcal{Z}}\cdot u_{\mathcal{Z}}%
}\left\vert \widetilde{x}_{\mathcal{A}}\right\rangle \otimes\\
\left\vert \phi_{u_{\mathcal{A}},u_{\mathcal{Z}},v_{\mathcal{X}}%
,u_{\mathcal{B}}}\right\rangle ^{B^{N}E^{N}}\left\vert u_{\mathcal{A}%
}\right\rangle \left\vert v_{\mathcal{X}}\right\rangle \left\vert
u_{\mathcal{B}}\right\rangle \left\vert \widetilde{x}_{\mathcal{B}%
}\right\rangle \left\vert u_{\mathcal{Z}}\right\rangle\otimes\ket{\widetilde{x}_{\mathcal{Z}}}.
\end{align*}
Thus, $\ket{\Psi_2'}$ is a superposition of polar encoded states as in (\ref{eq:phaseencstate}) and therefore the phase decoder will be useful to the receiver. In particular, Bob can first apply $U_{\mathcal{E}}^{\dagger C^N}$
and then apply
\[
\sum_{{x}_{\mathcal{A}},x_{\mathcal{Z}},{x}_{\mathcal{B}}%
}\sqrt{\Gamma_{{x}_{\mathcal{A}},x_{\mathcal{Z}}}^{\left(
{x}_{\mathcal{B}},u_{\mathcal{X}}\right)  }}\otimes\left\vert
\widetilde{x}_{\mathcal{A}}\right\rangle \left\vert \widetilde{x}_{\mathcal{Z}%
}\right\rangle \left\vert \widetilde{u}_{\mathcal{X}}\right\rangle \otimes\left\vert
\widetilde{x}_{\mathcal{B}}\right\rangle \left\langle \widetilde
{x}_{\mathcal{B}}\right\vert
\]
to coherently extract the values of ${x}_{\mathcal{A}}$ and $x_{\mathcal{Z}}$ using the frozen bits ${x}_{\mathcal{B}}$ and  ${u}_{\mathcal{X}}$. He then applies $U_{\mathcal{E}}^{C^N}$ to restore the $C^N$ registers to their previous form.  As with the amplitude decoding step, the closeness of the output of this process to the ideal output is governed by the error probability of the $W_P$ decoder (see Appendix~\ref{app:err-analysis}). 
To express the ideal output succinctly, we first make the assignments  
\begin{align*}
\left\vert \Phi_{\mathcal{A}}\right\rangle  &  \equiv\frac{1}{\sqrt
{2^{\left\vert \mathcal{A}\right\vert }}}\sum_{u_{\mathcal{A}}}\left\vert
u_{\mathcal{A}}\right\rangle \left\vert u_{\mathcal{A}}\right\rangle
,\ \left\vert \Phi_{\mathcal{Z}}\right\rangle \equiv\frac{1}{\sqrt
{2^{\left\vert \mathcal{Z}\right\vert }}}\sum_{v_{\mathcal{Z}}}\left\vert
v_{\mathcal{Z}}\right\rangle \left\vert v_{\mathcal{Z}}\right\rangle ,\\
\left\vert \Phi_{\mathcal{X}}\right\rangle  &  \equiv\frac{1}{\sqrt
{2^{\left\vert \mathcal{X}\right\vert }}}\sum_{v_{\mathcal{X}}}\left\vert
v_{\mathcal{X}}\right\rangle \left\vert v_{\mathcal{X}}\right\rangle
,\ \left\vert \Phi_{\mathcal{B}}\right\rangle \equiv\frac{1}{\sqrt
{2^{\left\vert \mathcal{B}\right\vert }}}\sum_{u_{\mathcal{B}}}\left\vert
u_{\mathcal{B}}\right\rangle \left\vert u_{\mathcal{B}}\right\rangle.
\end{align*}
Rewriting phase terms with Pauli operators, we then have that the actual output of this step of the decoder is $o(2^{-N^{\beta}})$-close in expected trace distance to the following ideal state:
\begin{multline*}
\ket{\Psi_3}=N_4\hspace{-3mm}\sum_{x_{\mathcal{A}},x_{\mathcal{B}},x_{\mathcal{Z}}}\hspace{-3mm}\ket{\widetilde{x}_{\mathcal{A}}}\otimes 
\ket{\widetilde{x}_{\mathcal{A}}}\ket{\widetilde{x}_{\mathcal{Z}}}\ket{\widetilde{u}_{\mathcal{X}}}\ket{\widetilde{x}_{\mathcal{B}}}\\
Z^{x_{\mathcal{A}},x_{\mathcal{Z}},u_{\mathcal{X}},x_{\mathcal{B}}}U_{\mathcal{N}}^{\otimes N}U_{\mathcal{E}}\ket{\Phi_{\mathcal{A}}} \ket{\Phi_{\mathcal{Z}}}\ket{\Phi_{\mathcal{X}}}\ket{\Phi_{\mathcal{B}}}\otimes \ket{\widetilde{x}_{\mathcal{Z}}},
\end{multline*}
where $N_4 \equiv 1/\sqrt{2^{|\mathcal{A}|+|\mathcal{B}|+|\mathcal{Z}|}}$.
Here $Z^{x_{\mathcal{A}},x_{\mathcal{Z}},u_{\mathcal{X}},x_{\mathcal{B}}}$ is shorthand for $Z^{x_{\mathcal{A}}}\otimes Z^{x_{\mathcal{Z}}}\otimes Z^{u_{\mathcal{X}}}\otimes Z^{x_{\mathcal{B}}}$, which acts on the second qubits in the entangled pairs, while the encoding and channel unitaries act on the first. 

The final step in the decoding process is to remove the phase operator $Z^{x_{\mathcal{A}},x_{\mathcal{Z}},v_{\mathcal{X}},x_{\mathcal{B}}}$ by controlled operations from the registers $\ket{\widetilde{x}_{\mathcal{A}}}\ket{\widetilde{x}_{\mathcal{Z}}}\ket{\widetilde{u}_{\mathcal{X}}}\ket{\widetilde{x}_{\mathcal{B}}}$ to the second qubits in the entangled pairs. This phase-basis controlled phase operation is equivalent to $N$ CNOT operations from the latter systems to the former and results in
\begin{multline*}%
\!\!\!\!N_0\sum_{x_{\mathcal{A}}}\ket{\widetilde{x}_{\mathcal{A}}}\otimes \ket{\widetilde{x}_{\mathcal{A}}} U_{\mathcal{N}}^{\otimes N}U_{\mathcal{E}}\ket{\Phi_{\mathcal{A},\mathcal{Z},\mathcal{X},\mathcal{B}}}
\sum_{x_{\mathcal{B}}}\ket{u_{\mathcal{X}}}\ket{\widetilde{x}_{\mathcal{B}}},
\end{multline*}
with Bob sharing $1/\sqrt{2^{|\mathcal{Z}|}}\sum_{x_{\mathcal{Z}}} \ket{\widetilde{x}_{\mathcal{Z}}}\otimes \ket{\widetilde{x}_{\mathcal{Z}}}$ with the inaccessible reference.
Thus the sender and receiver generate $|A|$ ebits with fidelity  $o(2^{-N^{\beta}})$ at the end of
the protocol.

\begin{remark}
The above scheme performs well with respect to a uniformly random choice of
the bits $u_{\mathcal{X}}$ and $u_{\mathcal{Z}}$, in the sense that the
expectation of the fidelity is high. Though, we can invoke Markov's inequality
to demonstrate that a large fraction of the possible codes have good performance.
\end{remark}

\begin{remark}
The first step of the decoder is identical to the first step of the decoder
from Ref.~\cite{WG11a}. Though, the second step above is an improvement over
the second step in Ref.~\cite{WG11a} because it is an explicit coherent
QSCD, rather than an inexplicit
controlled-decoupling unitary.
Additionally, the decoder's complexity is equivalent to $O\left(
N\right)  $\ quantum hypothesis tests and other unitaries resulting from the
 polar decompositions of
$\Lambda_{u_{\mathcal{A}},v_{\mathcal{X}}}^{\left(  u_{\mathcal{B}},u_{\mathcal{Z}}\right)}$ and
$\Gamma_{{x}_{\mathcal{A}},x_{\mathcal{Z}}}^{\left({x}_{\mathcal{B}},u_{\mathcal{X}}\right)}$,
but it remains unclear how to
implement these efficiently.
\end{remark}

\section{Zero e-bit rate for degradable channels}

\label{sec:ent-cons-rate-vanish}We can now prove that the entanglement consumption
rate of our quantum polar coding scheme vanishes for an arbitrary degradable
quantum channel.
We provide a brief summary of the proof (see Appendix~\ref{app-ent-cons-rate} for more
detail). Consider the following entropic
uncertainty principle \cite{RB09}:
$
H(X^{A}|B)_\rho  +H(  Z^{A}|E)_\rho  \geq 1,
$
where the conditional entropies are with respect to the phase and amplitude
observables $X$ and $Z$ measured with respect to a tripartite state
$\rho^{ABE}$ with $A$ being a qubit system. Using this and the fact that
$H(  X^A)  +H(  Z^A)  =2$ for our case, we can prove the
following uncertainty relation for the $i^{\text{th}}$ synthesized channels
$W_{P,N}^{\left(  i\right)  }$ and $W_{E,N}^{\left(  i\right)  }$:
$
I(W_{P,N}^{\left(  i\right)  })+I(W_{E,N}^{\left(  i\right)  })\leq1,
$
which is reminiscent of the relation $I\left(  X;BC\right)  =1-I\left(
Z;E\right)  $ mentioned previously. The above uncertainty relation then
implies the following one:
$
2\sqrt{F(  W_{P,N}^{\left(  i\right)  })  }+\sqrt{F(
W_{E,N}^{\left(  i\right)  })  }\geq 1.
$
This in turn implies that the phase-good channels to Bob are
amplitude-\textquotedblleft very bad\textquotedblright\ channels to Eve. From
degradability, we also know that the doubly-bad channels in
$\mathcal{B}$ are amplitude-bad channels to Eve. These two observations imply
that the phase-good channels to Bob, the doubly-bad channels to Bob, and
amplitude-good channels to Eve are disjoint sets. Furthermore, we know from
Theorem~\ref{thm:fraction-good}\ that the sum rate of the phase-good channels
to Bob and the amplitude-good channels to Eve is equal to $I\left(
X;BC\right)  +I\left(  Z;E\right)  =1-I\left(  Z;E\right)  +I\left(
Z;E\right)  =1$ as $N\rightarrow\infty$, implying that the rate of the
doubly-bad channel set $\mathcal{B}$ (the entanglement consumption rate) approaches zero in
the same limit. This same argument implies that the entanglement
consumption rate for the quantum polar codes in Ref.~\cite{WG11a}\ vanishes
for degradable quantum channels because the rate of the phase-good channels to
Bob is a lower bound on the rate of the amplitude-\textquotedblleft very
bad\textquotedblright\ channels to Eve.

\section{Superactivation}

\label{sec:superactivation}Our quantum polar coding scheme can be adapted to realize
the superactivation effect, in which two zero-capacity
quantum channels can \textit{activate} each other when used jointly, such that
the joint channel has a non-zero quantum capacity \cite{SY08}. Recall that the
channels from Ref.~\cite{SY08} are a four-dimensional PPT\ channel and a
four-dimensional 50\% erasure channel. Each of these have zero quantum
capacity, but the joint tensor-product channel has non-zero capacity.\footnote{We
are speaking of \textit{catalytic} superactivation. A
catalytic protocol  uses entanglement assistance, but the
figure of merit is the net rate of quantum communication---the total quantum
communication rate minus the entanglement consumption rate. Note
that the catalytic quantum capacity is equal to zero
if the standard quantum capacity is zero. Thus, the superactivation effect
that we speak of in this section is for the catalytic quantum capacity.}

We now discuss how to realize a quantum polar coding scheme for the joint
channel. Observe that the input space of the joint channel is 16-dimensional
and thus has a decomposition as a tensor product of four qubit-input
spaces:\ $\mathbb{C}^{4}\otimes\mathbb{C}^{4}\simeq\mathbb{C}^{16}%
\simeq\mathbb{C}^{2}\otimes\mathbb{C}^{2}\otimes\mathbb{C}^{2}\otimes
\mathbb{C}^{2}$. Thus, we can exploit a slightly modified version of our qubit
polar coding scheme. The idea is for Alice and Bob to employ a quantum polar
code for each qubit-input space in the tensor factor (this is similar to the
idea in Ref.~\cite{STA09}). There are amplitude and phase variables for each
of these qubit input spaces. Let $Z_{1}$, \ldots, $Z_{4}$ denote the amplitude
variables and let $X_{1}$, \ldots, $X_{4}$ denote the phase variables. Bob's
decoder is such that he coherently decodes $Z_{1}$, uses it as quantum side
information (QSI) to decode $Z_{2}$, uses both $Z_{1}$ and $Z_{2}$ as QSI
to decode $Z_{3}$, and then uses all of $Z_{1}$, \ldots, $Z_{3}$
to help decode $Z_{4}$. With all of the amplitude variables decoded, Bob then
uses these as QSI to decode $X_{1}$, and continues
successively until he coherently decodes $X_{4}$. At the end he performs
controlled phase gates to recover entanglement established with Alice.

We now calculate the total rate of this scheme. For the first qubit space in
the tensor factor, the channels split up into four types depending on whether
they are good/bad for amplitude/phase. Using the formula
(\ref{eq:net-rate-calc}) in Appendix~\ref{app:info-set-size}, the net quantum
data rate for the first tensor factor is equal to
$
I\left(  Z_{1};B\right)  +I\left(  X_{1};BZ_{1}Z_{2}Z_{3}Z_{4}\right)  -1.
$
(The formula is slightly different here because Bob decodes the phase variable
$X_{1}$ with all of the amplitude variables as QSI.) For
the second qubit space in the tensor factor, the net quantum data rate is
$
I\left(  Z_{2};BZ_{1}\right)  +I\left(  X_{2};BZ_{1}Z_{2}Z_{3}Z_{4}%
X_{1}\right)  -1.
$
We can similarly determine the respective net quantum data rates for the third
and fourth qubit spaces as%
$
I\left(  Z_{3};BZ_{1}Z_{2}\right)  +I\left(  X_{3};BZ_{1}Z_{2}Z_{3}
Z_{4}X_{1}X_{2}\right)  -1$,
$I\left(  Z_{4};BZ_{1}Z_{2}Z_{3}\right)  +I\left(  X_{4};BZ_{1}Z_{2}
Z_{3}Z_{4}X_{1}X_{2}X_{3}\right)  -1$.
Summing all these rates together with the chain rule and using the fact that
any two amplitude and/or phase variables are independent whenever $i\neq j$, we
obtain the overall net quantum data rate:
$
I\left(  Z_{1}Z_{2}Z_{3}Z_{4};B\right)  +I\left(  X_{1}X_{2}X_{3}X_{4}%
;BZ_{1}Z_{2}Z_{3}Z_{4}\right)  -4,
$
which is equal to the coherent information of the joint channel (by applying
the same Lemma~2 of Ref.~\cite{RB08}). The fact that our quantum
polar code can achieve the symmetric coherent information rate then proves
that superactivation occurs, given that Smith and Yard already showed that
this rate is non-zero for the channels mentioned above \cite{SY08}.

\section{Conclusion}

\label{sec:conclusion}Our quantum polar coding scheme has two benefits over
the work in Refs.~\cite{WG11a,RDR11}:\ it achieves the symmetric coherent
information rate for an arbitrary quantum channel and its entanglement
consumption rate vanishes for an arbitrary degradable channel. Though, we
should clarify that the analysis here actually implies that the scheme from
Ref.~\cite{WG11a}\ has the above two properties. A further benefit over the
scheme from Ref.~\cite{WG11a}\ is that the decoder here is explicitly realized
as $O\left(  N\right)  $ rounds of coherent quantum successive cancellation,
followed by $O\left(  N\right)  $ controlled-phase gates. Finally, we outlined how
the scheme here can exhibit the superactivation effect.
%
%
We acknowledge discussions with F.~Dupuis, S.~Guha, and
G.~Smith.

\bibliographystyle{IEEEtran}
\bibliography{Ref}

\onecolumn

\pagebreak
\appendices

\section{}

\label{app:info-set-size}We now calculate the rate of the set $\mathcal{A}$
(the rate of information qubits that Alice and Bob should be able to establish
with our quantum polar coding scheme). From basic set theory, we know that%
\begin{align*}
\left\vert \mathcal{A}\right\vert  &  =\left\vert \mathcal{G}_{N}\left(
W_{A},\beta\right)  \cap\mathcal{G}_{N}\left(  W_{P},\beta\right)  \right\vert
\\
&  =\left\vert \mathcal{G}_{N}\left(  W_{A},\beta\right)  \right\vert
+\left\vert \mathcal{G}_{N}\left(  W_{P},\beta\right)  \right\vert -\left\vert
\mathcal{G}_{N}\left(  W_{A},\beta\right)  \cup\mathcal{G}_{N}\left(
W_{P},\beta\right)  \right\vert .
\end{align*}
Given the polarization results for the cq amplitude and phase channels, we know that $\lim_{N\rightarrow\infty
}\frac{1}{N}\left\vert \mathcal{G}_{N}\left(  W_{A},\beta\right)  \right\vert
=I\left(  Z;B\right)  $ and $\lim_{N\rightarrow\infty}\frac{1}{N}\left\vert
\mathcal{G}_{N}\left(  W_{P},\beta\right)  \right\vert =I\left(  X;BC\right)
$. Also, consider that%
\begin{align*}
\left\vert \mathcal{G}_{N}\left(  W_{A},\beta\right)  \cup\mathcal{G}%
_{N}\left(  W_{P},\beta\right)  \right\vert  &  =\left\vert \left[  N\right]
\ \backslash\ \left(  \mathcal{G}_{N}\left(  W_{A},\beta\right)
\cup\mathcal{G}_{N}\left(  W_{P},\beta\right)  \right)  ^{c}\right\vert \\
&  =\left\vert \left[  N\right]  \ \backslash\ \left(  \mathcal{B}_{N}\left(
W_{A},\beta\right)  \cap\mathcal{B}_{N}\left(  W_{P},\beta\right)  \right)
\right\vert \\
&  =N-\left\vert \mathcal{B}\right\vert .
\end{align*}
Thus, the rate of $\mathcal{A}$ is equal to%
\begin{align}
\lim_{N\rightarrow\infty}\frac{1}{N}\left\vert \mathcal{A}\right\vert  &
=  I\left(  Z;B\right)  +I\left(  X;BC\right)  -1  +\lim
_{N\rightarrow\infty}\frac{1}{N}\left\vert \mathcal{B}\right\vert
\label{eq:net-rate-calc}\\
&  =  I\left(  Z;B\right)  +I\left(  X;BC\right)  -H\left(  Z\right)
  +\lim_{N\rightarrow\infty}\frac{1}{N}\left\vert \mathcal{B}%
\right\vert \nonumber\\
&  =I\left(  A\rangle B\right)  +\lim_{N\rightarrow\infty}\frac{1}%
{N}\left\vert \mathcal{B}\right\vert .\nonumber
\end{align}
where the second equality exploits the fact that $H\left(  Z\right)  =1$ for
a uniformly random bit and the third exploits Lemma~2 of Renes and Boileau
\cite{RB08}. Thus, the net rate of information qubits generated by this
quantum polar coding scheme is equal to the symmetric coherent information:%
\[
\lim_{N\rightarrow\infty}\frac{\left\vert \mathcal{A}\right\vert -\left\vert
\mathcal{B}\right\vert }{N}=I\left(  A\rangle B\right)  .
\]

\section{}

\label{app:err-analysis}We rigorously prove some of the
statements in Section~\ref{sec:err-analysis}. The ideal state after the first
step of the decoder is%
\[
\frac{1}{\sqrt{2^{\left\vert \mathcal{A}\right\vert +\left\vert \mathcal{B}%
\right\vert +\left\vert \mathcal{X}\right\vert }}}\sum_{u_{\mathcal{A}%
}^{\prime\prime},u_{\mathcal{B}}^{\prime\prime},v_{\mathcal{X}}^{\prime\prime
}}\left(  -1\right)  ^{u_{\mathcal{X}}\cdot v_{\mathcal{X}}^{\prime\prime}%
}\left\vert u_{\mathcal{A}}^{\prime\prime}\right\rangle \left\vert
\phi_{u_{\mathcal{A}}^{\prime\prime},u_{\mathcal{Z}},v_{\mathcal{X}}%
^{\prime\prime},u_{\mathcal{B}}^{\prime\prime}}\right\rangle ^{B^{N}E^{N}%
}\left\vert u_{\mathcal{A}}^{\prime\prime}\right\rangle \left\vert
v_{\mathcal{X}}^{\prime\prime}\right\rangle \left\vert u_{\mathcal{B}}%
^{\prime\prime}\right\rangle \left\vert u_{\mathcal{B}}^{\prime\prime
}\right\rangle \left\vert u_{\mathcal{Z}}\right\rangle .
\]
The actual state is%
\begin{align*}
&  \left(  \sum_{u_{\mathcal{A}}^{\prime},u_{\mathcal{B}}^{\prime
},v_{\mathcal{X}}^{\prime}}\sqrt{\Lambda_{u_{\mathcal{A}}^{\prime
},v_{\mathcal{X}}^{\prime}}^{\left(  u_{\mathcal{B}}^{\prime},u_{\mathcal{Z}%
}\right)  }}\otimes\left\vert u_{\mathcal{A}}^{\prime}\right\rangle \left\vert
v_{\mathcal{X}}^{\prime}\right\rangle \otimes\left\vert u_{\mathcal{B}%
}^{\prime}\right\rangle \left\vert u_{\mathcal{B}}^{\prime}\right\rangle
\left\langle u_{\mathcal{B}}^{\prime}\right\vert \otimes\left\vert
u_{\mathcal{Z}}\right\rangle \right)  \times\\
&  \ \ \ \ \ \ \left(  \frac{1}{\sqrt{2^{\left\vert \mathcal{A}\right\vert
+\left\vert \mathcal{B}\right\vert +\left\vert \mathcal{X}\right\vert }}}%
\sum_{u_{\mathcal{A}},u_{\mathcal{B}},v_{\mathcal{X}}}\left(  -1\right)
^{u_{\mathcal{X}}\cdot v_{\mathcal{X}}}\left\vert u_{\mathcal{A}}\right\rangle
\left\vert \phi_{u_{\mathcal{A}},u_{\mathcal{Z}},v_{\mathcal{X}}%
,u_{\mathcal{B}}}\right\rangle ^{B^{N}E^{N}}\left\vert u_{\mathcal{B}%
}\right\rangle \right)  \\
&  =\left(  \frac{1}{\sqrt{2^{\left\vert \mathcal{A}\right\vert +\left\vert
\mathcal{B}\right\vert +\left\vert \mathcal{X}\right\vert }}}\sum
_{\substack{u_{\mathcal{A}},u_{\mathcal{B}},v_{\mathcal{X}},\\u_{\mathcal{A}%
}^{\prime},v_{\mathcal{X}}^{\prime}}}\left(  -1\right)  ^{u_{\mathcal{X}}\cdot
v_{\mathcal{X}}}\left\vert u_{\mathcal{A}}\right\rangle \sqrt{\Lambda
_{u_{\mathcal{A}}^{\prime},v_{\mathcal{X}}^{\prime}}^{\left(  u_{\mathcal{B}%
},u_{\mathcal{Z}}\right)  }}\left\vert \phi_{u_{\mathcal{A}},u_{\mathcal{Z}%
},v_{\mathcal{X}},u_{\mathcal{B}}}\right\rangle ^{B^{N}E^{N}}\left\vert
u_{\mathcal{A}}^{\prime}\right\rangle \left\vert v_{\mathcal{X}}^{\prime
}\right\rangle \left\vert u_{\mathcal{B}}\right\rangle \left\vert
u_{\mathcal{B}}\right\rangle \left\vert u_{\mathcal{Z}}\right\rangle \right)
\end{align*}
The overlap between the above two states is equal to%
\begin{align*}
&  \frac{1}{2^{\left\vert \mathcal{A}\right\vert +\left\vert \mathcal{B}%
\right\vert +\left\vert \mathcal{X}\right\vert }}\sum
_{\substack{u_{\mathcal{A}}^{\prime\prime},u_{\mathcal{B}}^{\prime\prime
},v_{\mathcal{X}}^{\prime\prime},\\u_{\mathcal{A}},u_{\mathcal{B}%
},v_{\mathcal{X}},\\u_{\mathcal{A}}^{\prime},v_{\mathcal{X}}^{\prime}}}\left(
-1\right)  ^{u_{\mathcal{X}}\cdot\left(  v_{\mathcal{X}}^{\prime\prime
}+v_{\mathcal{X}}\right)  }\left\langle u_{\mathcal{A}}^{\prime\prime
}|u_{\mathcal{A}}\right\rangle \langle\phi_{u_{\mathcal{A}}^{\prime\prime
},u_{\mathcal{Z}},v_{\mathcal{X}}^{\prime\prime},u_{\mathcal{B}}^{\prime
\prime}}|\sqrt{\Lambda_{u_{\mathcal{A}}^{\prime},v_{\mathcal{X}}^{\prime}%
}^{\left(  u_{\mathcal{B}},u_{\mathcal{Z}}\right)  }}|\phi_{u_{\mathcal{A}%
},u_{\mathcal{Z}},v_{\mathcal{X}},u_{\mathcal{B}}}\rangle^{B^{N}E^{N}}\\
&  \ \ \ \ \ \ \ \ \ \ \ \ \ \ \ \ \ \ \ \ \ \ \ \ \ \ \ \ \times\left\langle
u_{\mathcal{A}}^{\prime}|u_{\mathcal{A}}^{\prime\prime}\right\rangle
\left\langle v_{\mathcal{X}}^{\prime}|v_{\mathcal{X}}^{\prime\prime
}\right\rangle \left\langle u_{\mathcal{B}}|u_{\mathcal{B}}^{\prime\prime
}\right\rangle \left\langle u_{\mathcal{B}}|u_{\mathcal{B}}^{\prime\prime
}\right\rangle \left\langle u_{\mathcal{Z}}|u_{\mathcal{Z}}\right\rangle \\
&  =\frac{1}{2^{\left\vert \mathcal{A}\right\vert +\left\vert \mathcal{B}%
\right\vert +\left\vert \mathcal{X}\right\vert }}\sum_{u_{\mathcal{A}%
},u_{\mathcal{B}},v_{\mathcal{X}},v_{\mathcal{X}}^{\prime}}\left(  -1\right)
^{u_{\mathcal{X}}\cdot\left(  v_{\mathcal{X}}^{\prime}+v_{\mathcal{X}}\right)
}\langle\phi_{u_{\mathcal{A}},u_{\mathcal{Z}},v_{\mathcal{X}}^{\prime
},u_{\mathcal{B}}}|\sqrt{\Lambda_{u_{\mathcal{A}},v_{\mathcal{X}}^{\prime}%
}^{\left(  u_{\mathcal{B}},u_{\mathcal{Z}}\right)  }}|\phi_{u_{\mathcal{A}%
},u_{\mathcal{Z}},v_{\mathcal{X}},u_{\mathcal{B}}}\rangle^{B^{N}E^{N}}%
\end{align*}
Taking the expectation of the fidelity over the uniformly random choice of the
ancilla bits $u_{\mathcal{X}}$ and $u_{\mathcal{Z}}$ then gives%
\begin{align*}
&  \mathbb{E}_{U_{\mathcal{X}},U_{\mathcal{Z}}}\left\{  \frac{1}{2^{\left\vert
\mathcal{A}\right\vert +\left\vert \mathcal{B}\right\vert +\left\vert
\mathcal{X}\right\vert }}\sum_{u_{\mathcal{A}},u_{\mathcal{B}},v_{\mathcal{X}%
},v_{\mathcal{X}}^{\prime}}\left(  -1\right)  ^{U_{\mathcal{X}}\cdot\left(
v_{\mathcal{X}}^{\prime}+v_{\mathcal{X}}\right)  }\langle\phi_{u_{\mathcal{A}%
},U_{\mathcal{Z}},v_{\mathcal{X}}^{\prime},u_{\mathcal{B}}}|\sqrt
{\Lambda_{u_{\mathcal{A}},v_{\mathcal{X}}^{\prime}}^{\left(  u_{\mathcal{B}%
},U_{\mathcal{Z}}\right)  }}|\phi_{u_{\mathcal{A}},U_{\mathcal{Z}%
},v_{\mathcal{X}},u_{\mathcal{B}}}\rangle^{B^{N}E^{N}}\right\}  \\
&  =\frac{1}{2^{\left\vert \mathcal{A}\right\vert +\left\vert \mathcal{B}%
\right\vert +\left\vert \mathcal{X}\right\vert +\left\vert \mathcal{Z}%
\right\vert }}\frac{1}{2^{\left\vert \mathcal{X}\right\vert }}\sum
_{u_{\mathcal{X}},u_{\mathcal{Z}}}\sum_{u_{\mathcal{A}},u_{\mathcal{B}%
},v_{\mathcal{X}},v_{\mathcal{X}}^{\prime}}\left(  -1\right)  ^{u_{\mathcal{X}%
}\cdot\left(  v_{\mathcal{X}}^{\prime}+v_{\mathcal{X}}\right)  }\langle
\phi_{u_{\mathcal{A}},u_{\mathcal{Z}},v_{\mathcal{X}}^{\prime},u_{\mathcal{B}%
}}|\sqrt{\Lambda_{u_{\mathcal{A}},v_{\mathcal{X}}^{\prime}}^{\left(
u_{\mathcal{B}},u_{\mathcal{Z}}\right)  }}|\phi_{u_{\mathcal{A}}%
,u_{\mathcal{Z}},v_{\mathcal{X}},u_{\mathcal{B}}}\rangle^{B^{N}E^{N}}\\
&  =\frac{1}{2^{\left\vert \mathcal{A}\right\vert +\left\vert \mathcal{B}%
\right\vert +\left\vert \mathcal{X}\right\vert +\left\vert \mathcal{Z}%
\right\vert }}\sum_{u_{\mathcal{A}},u_{\mathcal{B}},v_{\mathcal{X}%
},v_{\mathcal{X}}^{\prime},u_{\mathcal{Z}}}\left[  \frac{1}{2^{\left\vert
\mathcal{X}\right\vert }}\sum_{u_{\mathcal{X}}}\left(  -1\right)
^{u_{\mathcal{X}}\cdot\left(  v_{\mathcal{X}}^{\prime}+v_{\mathcal{X}}\right)
}\right]  \langle\phi_{u_{\mathcal{A}},u_{\mathcal{Z}},v_{\mathcal{X}}%
^{\prime},u_{\mathcal{B}}}|\sqrt{\Lambda_{u_{\mathcal{A}},v_{\mathcal{X}%
}^{\prime}}^{\left(  u_{\mathcal{B}},u_{\mathcal{Z}}\right)  }}|\phi
_{u_{\mathcal{A}},u_{\mathcal{Z}},v_{\mathcal{X}},u_{\mathcal{B}}}%
\rangle^{B^{N}E^{N}}\\
&  =\frac{1}{2^{\left\vert \mathcal{A}\right\vert +\left\vert \mathcal{B}%
\right\vert +\left\vert \mathcal{X}\right\vert +\left\vert \mathcal{Z}%
\right\vert }}\sum_{u_{\mathcal{A}},u_{\mathcal{B}},v_{\mathcal{X}%
},v_{\mathcal{X}}^{\prime},u_{\mathcal{Z}}}\delta_{v_{\mathcal{X}}^{\prime
},v_{\mathcal{X}}}\langle\phi_{u_{\mathcal{A}},u_{\mathcal{Z}},v_{\mathcal{X}%
}^{\prime},u_{\mathcal{B}}}|\sqrt{\Lambda_{u_{\mathcal{A}},v_{\mathcal{X}%
}^{\prime}}^{\left(  u_{\mathcal{B}},u_{\mathcal{Z}}\right)  }}|\phi
_{u_{\mathcal{A}},u_{\mathcal{Z}},v_{\mathcal{X}},u_{\mathcal{B}}}%
\rangle^{B^{N}E^{N}}\\
&  =\frac{1}{2^{\left\vert \mathcal{A}\right\vert +\left\vert \mathcal{B}%
\right\vert +\left\vert \mathcal{X}\right\vert +\left\vert \mathcal{Z}%
\right\vert }}\sum_{u_{\mathcal{A}},u_{\mathcal{B}},v_{\mathcal{X}%
},u_{\mathcal{Z}}}\langle\phi_{u_{\mathcal{A}},u_{\mathcal{Z}},v_{\mathcal{X}%
},u_{\mathcal{B}}}|\sqrt{\Lambda_{u_{\mathcal{A}},v_{\mathcal{X}}}^{\left(
u_{\mathcal{B}},u_{\mathcal{Z}}\right)  }}|\phi_{u_{\mathcal{A}}%
,u_{\mathcal{Z}},v_{\mathcal{X}},u_{\mathcal{B}}}\rangle^{B^{N}E^{N}}\\
&  \geq\frac{1}{2^{\left\vert \mathcal{A}\right\vert +\left\vert
\mathcal{B}\right\vert +\left\vert \mathcal{X}\right\vert +\left\vert
\mathcal{Z}\right\vert }}\sum_{u_{\mathcal{A}},u_{\mathcal{B}},v_{\mathcal{X}%
},u_{\mathcal{Z}}}\langle\phi_{u_{\mathcal{A}},u_{\mathcal{Z}},v_{\mathcal{X}%
},u_{\mathcal{B}}}|\Lambda_{u_{\mathcal{A}},v_{\mathcal{X}}}^{\left(
u_{\mathcal{B}},u_{\mathcal{Z}}\right)  }|\phi_{u_{\mathcal{A}},u_{\mathcal{Z}%
},v_{\mathcal{X}},u_{\mathcal{B}}}\rangle^{B^{N}E^{N}}\\
&  \geq1-o(2^{-N^{\beta}}),
\end{align*}
where the last inequality follows from the good performance of the quantum
successive cancellation decoder for the cq amplitude channels (see
Proposition~4 of Ref.~\cite{WG11}).

We can prove similarly that the phase decoder works well with a uniformly
random choice of the bits $u_{\mathcal{X}}$ and $u_{\mathcal{Z}}$. Observe
that a uniformly random choice of the bits $u_{\mathcal{Z}}$ induces a uniform
distribution of the bits $x_{\mathcal{Z}}$. A similar error analysis as above
then works for this case. Consider the ideal state:%
\[
\frac{1}{\sqrt{2^{\left\vert \mathcal{A}\right\vert +\left\vert \mathcal{B}%
\right\vert }}}\sum_{x_{\mathcal{A}},x_{\mathcal{B}}}\left\vert \widetilde
{x}_{\mathcal{A}}\right\rangle Z^{x_{\mathcal{A}},u_{\mathcal{X}%
},x_{\mathcal{B}},x_{\mathcal{Z}}}U_{\mathcal{N}}U_{\mathcal{E}}^{A^{\prime
N}}\left\vert \Phi_{\mathcal{A},\mathcal{Z},\mathcal{X},\mathcal{B}%
}\right\rangle \left\vert \widetilde{x}_{\mathcal{A}}\right\rangle \left\vert
\widetilde{x}_{\mathcal{Z}}\right\rangle \left\vert \widetilde{u}%
_{\mathcal{X}}\right\rangle \left\vert \widetilde{x}_{\mathcal{B}%
}\right\rangle ,
\]
and the actual state:%
\begin{align*}
&  U_{\mathcal{E}}^{C^{N}}\left(  \sum_{x_{\mathcal{A}}^{\prime}%
,x_{\mathcal{Z}}^{\prime},x_{\mathcal{B}}^{\prime}}\sqrt{\Gamma
_{x_{\mathcal{A}}^{\prime},x_{\mathcal{Z}}^{\prime}}^{\left(  x_{\mathcal{B}%
}^{\prime},u_{\mathcal{X}}\right)  }}\otimes\left\vert \widetilde
{x}_{\mathcal{A}}^{\prime}\right\rangle \left\vert \widetilde{x}_{\mathcal{Z}%
}^{\prime}\right\rangle \left\vert \widetilde{u}_{\mathcal{X}}\right\rangle
\otimes\left\vert \widetilde{x}_{\mathcal{B}}^{\prime}\right\rangle
\left\langle \widetilde{x}_{\mathcal{B}}^{\prime}\right\vert \right)
U_{\mathcal{E}}^{\dag C^{N}}\left(  \frac{1}{\sqrt{2^{\left\vert
\mathcal{A}\right\vert +\left\vert \mathcal{B}\right\vert }}}\sum
_{x_{\mathcal{A}},x_{\mathcal{B}}}\left\vert \widetilde{x}_{\mathcal{A}%
}\right\rangle Z^{x_{\mathcal{A}},u_{\mathcal{X}},x_{\mathcal{B}%
},x_{\mathcal{Z}}}U_{\mathcal{N}}U_{\mathcal{E}}^{A^{\prime N}}\left\vert
\Phi_{\mathcal{A},\mathcal{Z},\mathcal{X},\mathcal{B}}\right\rangle \left\vert
\widetilde{x}_{\mathcal{B}}\right\rangle \right)  \\
&  =\frac{1}{\sqrt{2^{\left\vert \mathcal{A}\right\vert +\left\vert
\mathcal{B}\right\vert }}}\sum_{x_{\mathcal{A}}^{\prime},x_{\mathcal{Z}%
}^{\prime},x_{\mathcal{B}}^{\prime},x_{\mathcal{A}}}\left\vert \widetilde
{x}_{\mathcal{A}}\right\rangle U_{\mathcal{E}}^{C^{N}}\sqrt{\Gamma
_{x_{\mathcal{A}}^{\prime},x_{\mathcal{Z}}^{\prime}}^{\left(  x_{\mathcal{B}%
},u_{\mathcal{X}}\right)  }}U_{\mathcal{E}}^{\dag C^{N}}Z^{x_{\mathcal{A}%
},u_{\mathcal{X}},x_{\mathcal{B}},x_{\mathcal{Z}}}U_{\mathcal{N}%
}U_{\mathcal{E}}^{A^{\prime N}}\left\vert \Phi_{\mathcal{A},\mathcal{Z}%
,\mathcal{X},\mathcal{B}}\right\rangle \left\vert \widetilde{x}_{\mathcal{B}%
}\right\rangle \left\vert \widetilde{x}_{\mathcal{A}}^{\prime}\right\rangle
\left\vert \widetilde{x}_{\mathcal{Z}}^{\prime}\right\rangle \left\vert
\widetilde{u}_{\mathcal{X}}\right\rangle .
\end{align*}
Now consider the overlap between the above two states:%
\begin{align*}
&  \frac{1}{2^{\left\vert \mathcal{A}\right\vert +\left\vert \mathcal{B}%
\right\vert }}\sum_{x_{\mathcal{A}}^{\prime\prime},x_{\mathcal{B}}%
^{\prime\prime}}\sum_{x_{\mathcal{A}}^{\prime},x_{\mathcal{Z}}^{\prime
},x_{\mathcal{B}}^{\prime},x_{\mathcal{A}}}\left\langle \widetilde
{x}_{\mathcal{A}}^{\prime\prime}|\widetilde{x}_{\mathcal{A}}\right\rangle
\times\\
&  \left\langle \Phi_{\mathcal{A},\mathcal{Z},\mathcal{X},\mathcal{B}%
}\right\vert U_{\mathcal{E}}^{\dag A^{\prime N}}U_{\mathcal{N}}^{\dag
}Z^{-x_{\mathcal{A}}^{\prime\prime},-u_{\mathcal{X}},-x_{\mathcal{B}}%
^{\prime\prime},-x_{\mathcal{Z}}}U_{\mathcal{E}}^{C^{N}}\sqrt{\Gamma
_{x_{\mathcal{A}}^{\prime},x_{\mathcal{Z}}^{\prime}}^{\left(  x_{\mathcal{B}%
},u_{\mathcal{X}}\right)  }}U_{\mathcal{E}}^{\dag C^{N}}Z^{x_{\mathcal{A}%
},u_{\mathcal{X}},x_{\mathcal{B}},x_{\mathcal{Z}}}U_{\mathcal{N}%
}U_{\mathcal{E}}^{A^{\prime N}}\left\vert \Phi_{\mathcal{A},\mathcal{Z}%
,\mathcal{X},\mathcal{B}}\right\rangle \times\\
&  \ \ \ \ \ \ \ \ \ \left\langle \widetilde{x}_{\mathcal{A}}^{\prime\prime
}|\widetilde{x}_{\mathcal{A}}^{\prime}\right\rangle \left\langle \widetilde
{x}_{\mathcal{Z}}|\widetilde{x}_{\mathcal{Z}}^{\prime}\right\rangle
\left\langle \widetilde{u}_{\mathcal{X}}|\widetilde{u}_{\mathcal{X}%
}\right\rangle \left\langle \widetilde{x}_{\mathcal{B}}^{\prime\prime
}|\widetilde{x}_{\mathcal{B}}\right\rangle \\
&  =\frac{1}{2^{\left\vert \mathcal{A}\right\vert +\left\vert \mathcal{B}%
\right\vert }}\sum_{x_{\mathcal{A}},x_{\mathcal{B}}}\left\langle
\Phi_{\mathcal{A},\mathcal{Z},\mathcal{X},\mathcal{B}}\right\vert
U_{\mathcal{E}}^{\dag A^{\prime N}}U_{\mathcal{N}}^{\dag}Z^{-x_{\mathcal{A}%
},-u_{\mathcal{X}},-x_{\mathcal{B}},-x_{\mathcal{Z}}}U_{\mathcal{E}}^{C^{N}%
}\sqrt{\Gamma_{x_{\mathcal{A}},x_{\mathcal{Z}}}^{\left(  x_{\mathcal{B}%
},u_{\mathcal{X}}\right)  }}U_{\mathcal{E}}^{\dag C^{N}}Z^{x_{\mathcal{A}%
},u_{\mathcal{X}},x_{\mathcal{B}},x_{\mathcal{Z}}}U_{\mathcal{N}%
}U_{\mathcal{E}}^{A^{\prime N}}\left\vert \Phi_{\mathcal{A},\mathcal{Z}%
,\mathcal{X},\mathcal{B}}\right\rangle \\
&  \geq\frac{1}{2^{\left\vert \mathcal{A}\right\vert +\left\vert
\mathcal{B}\right\vert }}\sum_{x_{\mathcal{A}},x_{\mathcal{B}}}\left\langle
\Phi_{\mathcal{A},\mathcal{Z},\mathcal{X},\mathcal{B}}\right\vert
U_{\mathcal{E}}^{\dag A^{\prime N}}U_{\mathcal{N}}^{\dag}Z^{-x_{\mathcal{A}%
},-u_{\mathcal{X}},-x_{\mathcal{B}},-x_{\mathcal{Z}}}U_{\mathcal{E}}^{C^{N}%
}\Gamma_{x_{\mathcal{A}},x_{\mathcal{Z}}}^{\left(  x_{\mathcal{B}%
},u_{\mathcal{X}}\right)  }U_{\mathcal{E}}^{\dag C^{N}}Z^{x_{\mathcal{A}%
},u_{\mathcal{X}},x_{\mathcal{B}},x_{\mathcal{Z}}}U_{\mathcal{N}%
}U_{\mathcal{E}}^{A^{\prime N}}\left\vert \Phi_{\mathcal{A},\mathcal{Z}%
,\mathcal{X},\mathcal{B}}\right\rangle
\end{align*}
Taking the expectation of this term over a uniformly random choice of
$u_{\mathcal{X}}$ and $u_{\mathcal{Z}}$ (which implies a uniformly random
choice of $x_{\mathcal{Z}}$) gives the following quantity:%
\begin{align*}
&  \mathbb{E}_{U_{\mathcal{X}},X_{\mathcal{Z}}}\left\{  \frac{1}{2^{\left\vert
\mathcal{A}\right\vert +\left\vert \mathcal{B}\right\vert }}\sum
_{x_{\mathcal{A}},x_{\mathcal{B}}}\left\langle \Phi_{\mathcal{A}%
,\mathcal{Z},\mathcal{X},\mathcal{B}}\right\vert U_{\mathcal{E}}^{\dag
A^{\prime N}}U_{\mathcal{N}}^{\dag}Z^{-x_{\mathcal{A}},-U_{\mathcal{X}%
},-x_{\mathcal{B}},-X_{\mathcal{Z}}}U_{\mathcal{E}}^{C^{N}}\Gamma
_{x_{\mathcal{A}},X_{\mathcal{Z}}}^{\left(  x_{\mathcal{B}},U_{\mathcal{X}%
}\right)  }U_{\mathcal{E}}^{\dag C^{N}}Z^{x_{\mathcal{A}},U_{\mathcal{X}%
},x_{\mathcal{B}},X_{\mathcal{Z}}}U_{\mathcal{N}}U_{\mathcal{E}}^{A^{\prime
N}}\left\vert \Phi_{\mathcal{A},\mathcal{Z},\mathcal{X},\mathcal{B}%
}\right\rangle \right\}  \\
&  =\frac{1}{2^{\left\vert \mathcal{A}\right\vert +\left\vert \mathcal{B}%
\right\vert +\left\vert \mathcal{X}\right\vert +\left\vert \mathcal{Z}%
\right\vert }}\sum_{x_{\mathcal{A}},x_{\mathcal{B}},u_{\mathcal{X}%
},x_{\mathcal{Z}}}\left\langle \Phi_{\mathcal{A},\mathcal{Z},\mathcal{X}%
,\mathcal{B}}\right\vert U_{\mathcal{E}}^{\dag A^{\prime N}}U_{\mathcal{N}%
}^{\dag}Z^{-x_{\mathcal{A}},-u_{\mathcal{X}},-x_{\mathcal{B}},-x_{\mathcal{Z}%
}}U_{\mathcal{E}}^{C^{N}}\Gamma_{x_{\mathcal{A}},x_{\mathcal{Z}}}^{\left(
x_{\mathcal{B}},u_{\mathcal{X}}\right)  }U_{\mathcal{E}}^{\dag C^{N}%
}Z^{x_{\mathcal{A}},u_{\mathcal{X}},x_{\mathcal{B}},x_{\mathcal{Z}}%
}U_{\mathcal{N}}U_{\mathcal{E}}^{A^{\prime N}}\left\vert \Phi_{\mathcal{A}%
,\mathcal{Z},\mathcal{X},\mathcal{B}}\right\rangle \\
&  \geq1-o(2^{-N^{\beta}}),
\end{align*}
where the last inequality again follows from the performance of the quantum
successive cancellation decoder for the phase channels.

\section{}

\label{app-ent-cons-rate}This appendix provides a detailed proof that the
entanglement consumption rate of our quantum polar codes vanishes whenever the
quantum channel is degradable. Consider a state of the following form:%
\[
\frac{1}{\sqrt{2^N}} \sum_{z^{N}}\left\vert z^{N}\right\rangle ^{A^{N}}\left\vert \phi_{z^{N}G_N%
}\right\rangle ^{B^{N}E^{N}}\left\vert z^{N}\right\rangle ^{C^{N}}.
\]
We can represent the registers $A_{1}\cdots A_{i-1}$ in the amplitude basis
and the registers $A_{i+1}\cdots A_{N}$ in the phase basis as follows:%
\[
\frac{1}{\sqrt{2^{2N-i}}} \sum_{z^{N},x^N_{i+1}}\left\vert z_{1}^{i-1}\right\rangle ^{A_{1}^{i-1}}\left\vert
z_{i}\right\rangle ^{A_{i}}\left\vert \widetilde{x}_{i+1}^{N}\right\rangle ^{A_{i+1}^{N}%
}\left\vert \phi_{z^{N}G_N}\right\rangle ^{B^{N}E^{N}}\left(  Z^{x_{i+1}^{N}%
}\right)  ^{C_{i+1}^{N}}\left\vert z^{N}\right\rangle ^{C^{N}}.
\]
Then measuring the systems $A_{1}\cdots A_{i-1}$ in the amplitude basis and
the systems $A_{i+1}\cdots A_{N}$ in the phase basis (we can think of this
just as dephasing these systems in the respective bases) leads to a state
which can generate the outputs of  the
$i^{\text{th}}$ phase channel to Bob $W_{P,N}^{\left(  i\right)  }$ and the
$i^{\text{th}}$ amplitude channel to Eve $W_{P,N}^{\left(  i\right)  }$. Denote this state by $\psi_i$, and call
the various measurement outputs systems $Z_{1}\cdots Z_{i-1}$ and $X_{i+1}\cdots X_{N}$ in order
to indicate that they are classical. Then $\psi_i$ is a tripartite state on  $A_{i}|B^{N}C^{N}X_{i+1}^{N}|E^{N}%
Z_{1}^{i-1}$ is a tripartite state (where the vertical bars indicate the
divisions of the parties), to which we can apply the following uncertainty relation proved in Ref.~\cite{RB09}:%
\[
H\left(  X^{A_{i}}|B^{N}C^{N}X_{i+1}^{N}\right)_{\psi_i}  +H\left(  Z^{A_{i}}%
|E^{N}Z_{1}^{i-1}\right)_{\psi_i}  \geq1.
\]
Combining this with $H\left(  X^{A_{i}}\right)
+H\left(  Z^{A_{i}}\right)  =2$ (which holds because $X^{A_{i}}$ and
$Z^{A_{i}}$ are uniform random bits) gives%
\[
I\left(  X^{A_{i}};B^{N}C^{N}X_{i+1}^{N}\right)_{\psi_i} +I\left(  Z^{A_{i}}%
;E^{N}Z_{1}^{i-1}\right)_{\psi_i}  \leq1,
\]
or equivalently,%
\begin{equation}
I(W_{P,N}^{\left(  i\right)  })+I(W_{E,N}^{\left(  i\right)  })\leq
1.\label{eq:channel-uncertainty-relation}%
\end{equation}

Note that in the limit $N\rightarrow \infty$, the channels polarize, so that the channels which are
good in phase for Bob are bad in amplitude for Eve, and the ones which are
good in amplitude for Eve are bad in phase for Bob. This demonstrates that our
quantum polar coding scheme given here is asymptotically equivalent to the
scheme of Wilde and Guha \cite{WG11a}\ in the limit of many recursions of the encoding
after the channel polarization effect takes hold.

The above uncertainty relation is helpful in proving a different one about the
synthesized channels' fidelities, that will in turn help us prove the
statement about the entanglement consumption rate. Exploiting the following
inequality (see Proposition~1 of Ref.~\cite{WG11})%
\begin{equation}
I\left(  W\right)  \geq\log_{2}\left(  \frac{2}{1+\sqrt{F\left(  W\right)  }%
}\right)  ,\label{eq:arikan-inequality}%
\end{equation}
we can show that%
\begin{align}
2\sqrt{F\left(  W_{P,N}^{\left(  i\right)  }\right)  }+\sqrt{F\left(
W_{E,N}^{\left(  i\right)  }\right)  } &  \geq
1,\label{eq:fidelity-uncertainty-relations}\\
\sqrt{F\left(  W_{P,N}^{\left(  i\right)  }\right)  }+2\sqrt{F\left(
W_{E,N}^{\left(  i\right)  }\right)  } &  \geq1.\nonumber
\end{align}
Consider that the inequality in (\ref{eq:arikan-inequality}) above is
equivalent to
$
\sqrt{F\left(  W\right)  }\geq2^{1-I\left(  W\right)  }-1.
$
We then have%
\[
\sqrt{F\left(  W_{P,N}^{\left(  i\right)  }\right)  }\ \ \ \geq
\ \ \ 2^{1-I\left(  W_{P,N}^{\left(  i\right)  }\right)  }-1\ \ \ \geq
\ \ \ 2^{I\left(  W_{E,N}^{\left(  i\right)  }\right)  }-1\ \ \ \geq
\ \ \ \frac{2}{1+\sqrt{F\left(  W_{E,N}^{\left(  i\right)  }\right)  }}-1,
\]
where we used the uncertainty relation in
(\ref{eq:channel-uncertainty-relation}) in the second inequality and we again
applied (\ref{eq:arikan-inequality}) for the third inequality. Rewriting this,
we obtain%
\[
\left(  1+\sqrt{F\left(  W_{E,N}^{\left(  i\right)  }\right)  }\right)
\sqrt{F\left(  W_{P,N}^{\left(  i\right)  }\right)  }\geq2-\left(
1+\sqrt{F\left(  W_{E,N}^{\left(  i\right)  }\right)  }\right)  ,
\]
which gives%
\[
2\sqrt{F\left(  W_{P,N}^{\left(  i\right)  }\right)  }+\sqrt{F\left(
W_{E,N}^{\left(  i\right)  }\right)  }\geq1.
\]
(We used the fact that the fidelity is less than one.)\ Proceeding
in the symmetric way gives the other fidelity uncertainty relation in
(\ref{eq:fidelity-uncertainty-relations}).

We can now argue that the entanglement consumption rate should be zero in the
limit whenever the channel is a degradable quantum channel. We do this by a
modification of the argument in Ref.~\cite{WG11a}. Consider the set of
channels $\mathcal{B}$ which are doubly bad for amplitude and phase. Also,
consider the channels which are amplitude-good for Eve:%
\[
\mathcal{G}_{N}\left(  W_{E},\beta\right)  \equiv\left\{  i\in\left[
N\right]  :\sqrt{F(W_{E,N}^{\left(  i\right)  })}<2^{-N^{\beta}}\right\}
\]
and those which are phase-good for Bob: $\mathcal{G}_{N}\left(  W_{P}%
,\beta\right)  $. We prove now that these sets are disjoint and thus the sum
of them must be smaller than $N$ (the total number of channel uses). First
consider that%
\[
\mathcal{B\cap G}_{N}\left(  W_{P},\beta\right)  =\emptyset
\]
by the definition of the set $\mathcal{B}$. Now consider that%
\[
2\cdot2^{-N^{\beta}}\geq2\cdot\sqrt{F(W_{P,N}^{\left(  i\right)  })}%
\geq1-\sqrt{F(W_{E,N}^{\left(  i\right)  })},
\]
implying that%
\[
\sqrt{F(W_{E,N}^{\left(  i\right)  })}\geq1-2\cdot2^{-N^{\beta}},
\]
whenever $2^{-N^{\beta}}\geq\sqrt{F(W_{P,N}^{\left(  i\right)  })}$.
Thus, all of the channels that are phase-good for Bob are
amplitude-\textquotedblleft very bad\textquotedblright\ for Eve. So the
following relation holds for large enough $N$:%
\[
\mathcal{G}_{N}\left(  W_{P},\beta\right)  \cap\mathcal{G}_{N}\left(
W_{E},\beta\right)  =\emptyset.
\]
The relation
$
\mathcal{B}\cap\mathcal{G}_{N}\left(  W_{E},\beta\right)  =\emptyset
$
holds for degradable channels because%
\begin{align*}
&  \mathcal{B}\cap\mathcal{G}_{N}\left(  W_{E},\beta\right)  \\
&  =\left(  \mathcal{B}_{N}\left(  W_{A},\beta\right)  \cap\mathcal{B}%
_{N}\left(  W_{P},\beta\right)  \right)  \cap\mathcal{G}_{N}\left(
W_{E},\beta\right)  \\
&  \subseteq\left(  \mathcal{B}_{N}\left(  W_{E},\beta\right)  \cap
\mathcal{B}_{N}\left(  W_{P},\beta\right)  \right)  \cap\mathcal{G}_{N}\left(
W_{E},\beta\right)  \\
&  =\emptyset.
\end{align*}
The second line follows from the definition and the third follows from the
degradability condition (all the channels that are bad in amplitude for Bob
are also bad in amplitude for Eve due to the existence of a degrading map
under which the fidelity can only increase---see Lemma~3 of Ref.~\cite{WG11a}%
). Thus, all of these sets are disjoint and it follows that%
\[
\frac{1}{N}\left(  \left\vert \mathcal{G}_{N}\left(  W_{E},\beta\right)
\right\vert +\left\vert \mathcal{G}_{N}\left(  W_{P},\beta\right)  \right\vert
+\left\vert \mathcal{B}\right\vert \right)  \leq1.
\]

Finally, we know from Theorem~\ref{thm:fraction-good} that the rates of the
sets $\mathcal{G}_{N}\left(  W_{E},\beta\right)  $ and $\mathcal{G}_{N}\left(
W_{P},\beta\right)  $ in the asymptotic limit are%
\begin{align*}
\lim_{N\rightarrow\infty}\frac{1}{N}\left\vert \mathcal{G}_{N}\left(
W_{E},\beta\right)  \right\vert  &  =I\left(  Z^A;E\right)_\psi  ,\\
\lim_{N\rightarrow\infty}\frac{1}{N}\left\vert \mathcal{G}_{N}\left(
W_{P},\beta\right)  \right\vert  &  =I\left(  X^A;BC\right)_\psi  =1-I\left(
Z^A;E\right)_\psi  ,
\end{align*}
so that the rate of $\mathcal{B}$ must be zero in the asymptotic limit:%
\[
\lim_{N\rightarrow\infty}\frac{1}{N}\left\vert \mathcal{B}\right\vert =0.
\]

\end{document}